\documentclass[aps,floats,floatfix,showpacs,onecolumn,twoside,preprintnumbers,superscriptaddress,nofootinbib]{revtex4}
\usepackage{amssymb,amsmath,pstricks}
\usepackage[dvips]{graphicx}

\def\openone{\leavevmode\hbox{\small1\kern-3.8pt\normalsize1}}%
\def\sig{\mbox{Sg}\,}

\def\bea{\begin{eqnarray}}
\def\eea{\end{eqnarray}}
\def\beq{\begin{equation}}
\def\eeq{\end{equation}}

\begin{document}

\title{Color neutrality effects in the phase diagram of the PNJL model}
\author{D.~G\'omez Dumm}
\email{dumm@fisica.unlp.edu.ar}
\affiliation{IFLP, CONICET $-$ Dpto.\ de F\'{\i}sica, Universidad Nacional de
La Plata, C.C. 67, 1900 La Plata, Argentina}
\affiliation{CONICET, Rivadavia 1917, 1033 Buenos Aires, Argentina}
\author{D.~B.~Blaschke}
\email{blaschke@ift.uni.wroc.pl}
\affiliation{Institute for Theoretical Physics, University of Wroc{\l}aw,
Max Born place 9, 50204 Wroc{\l}aw, Poland}
\affiliation{Bogoliubov  Laboratory of Theoretical Physics, JINR Dubna,
Joliot-Curie Street 6, 141980  Dubna, Russia}
\author{A.G. Grunfeld\footnote{Present address: Institute for Nuclear Physics,
TU Darmstadt, Schlo{\ss}gartenstr. 9, D-64289 Darmstadt, Germany \\ \hfill }}
\email{ag.grunfeld@gmail.com}
\affiliation{CONICET, Rivadavia 1917, 1033 Buenos Aires, Argentina}
\affiliation{Physics Department, Comisi\'on Nacional de
Energ\'{\i}a At\'omica, Av.\ Libertador 8250, 1429 Buenos Aires, Argentina}
\author{N. N. Scoccola}
\email{scoccola@tandar.cnea.gov.ar}
\affiliation{CONICET, Rivadavia 1917, 1033 Buenos Aires, Argentina}
\affiliation{Physics Department, Comisi\'on Nacional de
Energ\'{\i}a At\'omica, Av.\ Libertador 8250, 1429 Buenos Aires, Argentina}
\affiliation{Universidad Favaloro, Sol{\'\i}s 453, 1078 Buenos Aires,
Argentina}
\date{\today}

\begin{abstract}
The phase diagram of a two-flavor Polyakov loop Nambu$-$Jona-Lasinio model
is analyzed imposing the constraint of color charge neutrality. The main
effect of this constraint is a coexistence of the chiral symmetry breaking
($\chi$SB) and two-flavor superconducting phases. Additional effects are a
shrinking of the $\chi$SB domain in the $T-\mu$ plane and a shift of the
end point to lower temperatures, but their quantitative importance is
shadowed by the intrinsic uncertainties of the model. The effects can be
understood in view of the presence of a nonvanishing color chemical
potential $\mu_8$, which is introduced to compensate the color charge
density $\rho_8$ induced by a background color gauge mean field $\phi_3$.
At low temperatures and large chemical potentials the model exhibits a
quarkyonic phase, which gets additional support from the diquark
condensation.
\end{abstract}


\maketitle

\renewcommand{\thefootnote}{\arabic{footnote}}
\setcounter{footnote}{0}

\section{Introduction}

Constraints on the phase diagram of QCD under extreme conditions of high
excitation (temperature) and compression (density) are of vital interest
for large-scale experimental programmes with ultrarelativistic heavy-ion
beams, searching for signatures of the QCD phase transitions: chiral
symmetry restoration and deconfinement.

Under conditions of finite temperature $T$ and small chemical potentials
$\mu$, which are probed in heavy ion collisions at SPS, RHIC and in the
near future at LHC, Lattice QCD simulations have provided insight into the
phase structure since the temperature-dependent change of the order
parameters of the above phases, namely the chiral condensate $\langle \bar
qq\rangle$ and the traced Polyakov loop $\Phi$, have been
determined~\cite{Karsch,Karsch:2006xs,Cheng:2006qk,Cheng:2007jq}. The peak
positions of the susceptibilities derived from these calculations for both
transitions are remarkably coincident at the same pseudocritical
temperature $T_c$. This might be understood in terms of general effective
theory~\cite{Mocsy:2003qw}. It is also remarkable that these results can
be nicely described with an effective model of the NJL type, extended with
the inclusion of a coupling to the Polyakov loop
\cite{Fukushima:2003ib,Fukushima:2003fw,Fukushima:2003fm,Fukushima:2002ew}
with a  potential ${\cal U}(\Phi,T)$ that can be extracted from Lattice
QCD simulations of the pressure in the pure gauge theory
\cite{Ratti:2005jh,Roessner:2006xn}.

The challenge for experiments as well as for theory is to extend this
knowledge into the domain of finite baryon densities, where precursors of
color superconductivity (pseudogap phase~\cite{kitazawa}) or even color
superconducting quark matter phases themselves can
occur~\cite{buballa,alford} and new constraints from observations of
compact star properties may apply \cite{Klahn:2006iw,Alford:2006vz}.

Most of the effective model analyses addressing the chiral phase transition
in the QCD phase diagram have been performed with NJL type models which,
however, lack quark confinement and lead to an onset of finite quark densities
already at unphysically low temperatures of about 50 MeV \cite{Zhuang:1994dw},
far below the deconfinement temperature of about 200 MeV
\cite{Cheng:2006qk,Cheng:2007jq}.

The straightforward step to implement the effects of confining forces and
to suppress unphysical quark degrees of freedom is to study the phase
diagram of quark matter in the Polyakov loop NJL (PNJL) model. This
has been carried out, e.g., in
Refs.~\cite{Roessner:2006xn,Sasaki:2006ww,Rossner:2007ik,Fukushima:2008wg,
Ciminale:2007sr,Abuki:2008nm}. In these works, however, the important
question of color neutrality of quark matter has not been considered. The
inclusion of neutrality constraints in the PNJL model has been taken up
recently in Ref.~\cite{Abuki:2008ht}. In that paper, however, only the
high-density region for a quark chemical potential $\mu=500$~MeV has been
considered, and light quark masses have been neglected. It has been found
that the Polyakov-loop formulation together with neutrality constraints
can give rise to an increase of the critical temperature for the color
superconductivity transition, in deviation from the well-known BCS
relationship. An investigation of the interrelation between chiral
symmetry breaking ($\chi$SB) and color superconductivity, as well as a
study of the effect of color neutrality without enforcing electric
neutrality, is still missing. The aim of the present paper is to address
these subjects, analyzing the QCD phase diagram in the framework of a
generalization of the PNJL model that fulfills the constraint of color
neutrality.

The article is organized as follows. In Section II we introduce the
formalism for enforcing color neutrality in the PNJL model by color
chemical potentials. In Section III we discuss the numerical results for
the phase diagram, first in the non-superconducting case, then including
color superconductivity. In the final section we present our conclusions
and discuss possible consequences for the phenomenology of next heavy-ion
collision experiments.

\section{Formalism}

The Euclidean action for the two-flavor PNJL at temperature $T$ is
given by
\begin{eqnarray}
S_E & = & \int_0^\beta d\tau \int d^3 x \;\bigg\{ \bar \psi \left(
-\,i \gamma_\mu D_\mu + \hat m \right) \psi \; - \; \frac{G}{2}\, \left[
(\bar\psi \psi)^2 \, + \, (\bar\psi
i\gamma_5\vec\tau \psi)^2\right] \nonumber \\
& & \ - \; \frac{H}{2}\, (\bar \psi_C \, i \gamma_5
\tau_2 \lambda_a \, \psi) \, (\bar \psi_C \, i \gamma_5 \tau_2 \lambda_a
\, \psi)^\dagger \; + \; {\cal U}(\Phi,T) \bigg\}\;,
\label{se}
\end{eqnarray}
where $\psi$ is the $N_f=2$ fermion doublet $\psi \equiv (u\; d)^T$, $\hat
m = {\rm diag}(m_u, m_d)$ stands for the current quark mass matrix,
and $\beta=1/T$. For simplicity we consider the isospin symmetry limit,
in which $m_u = m_d=\bar m_0$. Charge conjugated fields in Eq.~(\ref{se})
are defined by $\psi_C = \gamma_2\gamma_4 \,\bar \psi^T$, while $\vec
\tau$ and $\lambda_a$, with $a=2,5,7$, stand for Pauli and Gell-Mann
matrices acting on flavor and color spaces, respectively. This Euclidean
action leads to local chiral invariant current-current interactions in the
quark-antiquark and quark-quark channels. The latter is expected to be
responsible for the presence of a color superconducting phase in the
region of low temperatures and moderate chemical potentials.

The coupling of fermions to the Polyakov loop has been implemented in
Eq.~(\ref{se}) through the covariant derivative in the fermion kinetic
term $\gamma_\mu D_\mu$, where $D_\mu\equiv \partial_\mu - iA_\mu$. Here
the Euclidean operator $\gamma_\mu\partial_\mu$ is defined as
$\gamma_4\frac{\partial}{\partial \tau} + \vec\gamma\cdot \vec\nabla$,
with $\gamma_4=i\gamma_0$. As usual, we assume that the quarks move in a
background gauge field $A_0 = g\,\delta_{\mu 0}\, G^\mu_a \lambda^a/2$,
where $G^\mu_a$ are the SU(3) color gauge fields. Then, at the mean field
level, the traced Polyakov loop is given by $\Phi=\frac{1}{3} {\rm Tr}\,
\exp( i\beta \phi)$, with $\phi = i\bar A_0= {\rm constant}$. The traced
Polyakov loop can be taken as an order parameter for the confinement
transition. In the pure glue theory it can be associated with the
spontaneous breaking of the global $Z_3$ center symmetry of color $SU(3)$,
$\Phi=0$ corresponding to the symmetric, confined
phase~\cite{Pisarski:2000eq}. Here we will work in the so-called Polyakov
gauge, in which the matrix $\phi$ is given a diagonal representation $\phi
= \phi_3 \lambda_3 + \phi_8 \lambda_8$, which leaves only two independent
variables, $\phi_3$ and $\phi_8$. Finally, the action (\ref{se}) also
includes an effective potential ${\cal U}$ that accounts for gauge field
self-interactions.

In order to obtain the grand canonical thermodynamical potential
of the PNJL model at temperature $T$ and chemical potential
$\mu=\mu_B/3$ we start from Eq.(\ref{se}), performing a standard
bosonization of the theory. This can be done by introducing fields
$(\sigma, \vec \pi)$ and $\Delta_a$ corresponding to the
quark-antiquark and diquark channels respectively. Then we
consider the mean field approximation, keeping the nonzero vacuum
expectation values $\bar\sigma$ and $\phi$ and dropping the
respective fluctuations. Concerning the superconducting vacuum, we
adopt here the usual ansatz in which one has $\bar \Delta_5 =\bar
\Delta_7=0$, $\bar \Delta_2= \bar \Delta$. In this way, using the
standard Matsubara formalism one obtains
\begin{equation}
\Omega_{\rm MFA} \ = \ - \, \frac{T}{2} \sum_{n=-\infty}^\infty \int
\frac{d^3p}{(2\pi)^3}\ \ln \; \det \left[ \beta\, S^{-1}(T,\mu)\right] + \;
\frac{\bar\sigma^2}{2G}\; + \; \frac{\bar\Delta^2}{2H} \; + \;
{\cal{U}}(\Phi ,T) \ , \label{ommfaI}
\end{equation}
where the inverse propagator $S^{-1}$ is a $48 \times 48$ matrix in Dirac,
flavor, color and Nambu-Gorkov spaces. The dependence of $S^{-1}$ with $T$
and $\mu$ ($\mu=\mu_B/3$) is obtained from the $T=\mu=0$ four-momentum
integrals in the fermion determinant by replacing $p_4 \to \omega_n -
i\tilde\mu$, where $\omega_n = (2n+1)\pi T$ are the usual fermionic
Matsubara frequencies, and $\tilde\mu$ stands for a complex ``chemical
potential'' matrix in color space,
\begin{equation}
\tilde\mu \ = \ {\rm diag}(\mu_r,\mu_g,\mu_b) \ + \
i\, (\phi_3 \, \lambda_3 \; + \; \phi_8\, \lambda_8) \ ,
\end{equation}
with
\begin{equation}
{\rm diag}(\mu_r,\mu_g,\mu_b) \ = \ \mu\;
\openone\; +\; \mu_3 \, \lambda_3 \; + \; \mu_8 \, \lambda_8 \ .
\end{equation}
In general, the inclusion of color chemical potentials $\mu_3$ and $\mu_8$
is necessary to ensure color neutrality. Concerning the mean field
effective potential ${\cal{U}}(\Phi ,T)$, which accounts for the
Polyakov loop dynamics, we use here a form that appears to be consistent
with group theory constraints as well as lattice results (from which one
can estimate the temperature dependence). Following
Ref.~\cite{Roessner:2006xn} we take the logarithmic form of the
Polyakov-loop potential, motivated by the SU(3) Haar measure
\begin{equation}
\frac{{\cal U}(\Phi , T)}{T^4} \ = \ -\,\frac{1}{2}\,
a(T)\,\Phi^\ast\Phi \; + \; b(T)\, \ln \left[\, 1 - 6\, \Phi^\ast\Phi +
4\, (\Phi^3 + \Phi^{\ast 3}) - 3\, (\Phi^\ast\Phi)^2\, \right] \ ,
\end{equation}
with the corresponding definitions of $a(T)$ and $b(T)$ from
\cite{Roessner:2006xn}.

It is worth to notice that in presence of the Polyakov loop the
thermodynamical potential could be in general a complex quantity. As
discussed in Ref.~\cite{Rossner:2007ik}, in order to properly define the
mean field approximation one should require that the mean field
configuration provides the maximal contribution to the partition function.
Then, the mean field values (order parameters) $\bar\sigma$, $\bar\Delta$,
$\phi_3$ and $\phi_8$ should fulfill the coupled set of ``gap
equations''~\cite{Rossner:2007ik}
\begin{equation}
\frac{\partial {\rm Re}\, [\Omega_{\rm MFA}]}
{\left( \partial\bar\sigma, \partial\bar\Delta, \partial\phi_3,
\partial\phi_8 \right)} \ = \ 0.
\label{gap}
\end{equation}
If one also imposes color charge neutrality conditions, one has two
additional equations that fix the color chemical potentials $\mu_3$ and
$\mu_8$. The vanishing of color charges implies
\begin{equation}
\,\frac{\partial {\rm Re}\,
[\Omega_{\rm MFA}]}{\left( \partial\mu_3, \partial\mu_8 \right)}\ =\ 0.
\label{charges}
\end{equation}
Finally, in order to obtain a physically meaningful solution, it has to be
verified that the resulting field configuration leads to a real-valued
thermodynamical potential.

At vanishing chemical potential, owing to the charge conjugation
properties of the QCD Lagrangian, the mean field traced Polyakov loop
$\Phi$ is expected to be a real quantity. Since $\phi_3$ and $\phi_8$ have
to be real-valued~\cite{Rossner:2007ik}, this condition implies $\phi_8 =
0$, and it is easy to see that the thermodynamical potential turns out to
be real, as desired. In general, this need not to be the case at finite
$\mu$ \cite{Elze:1987,Dumitru:2005ng,Fukushima:2006uv}. As in
Refs.\cite{Roessner:2006xn,Abuki:2008ht,Rossner:2007ik} we will assume
that the potential $\cal U$ is such that the condition $\phi_8=0$ is well
satisfied for the range of values of $\mu$ and $T$ investigated here. The
traced Polyakov loop is then given by $\Phi = \Phi^\ast = [ 1 + 2\,\cos
(\beta\phi_3) ]/3$. In addition, it can be seen that if $\phi_8 = 0$ then
Eqs.~(\ref{gap}) and (\ref{charges}) lead to $\mu_3 = 0$, leaving $\mu_8$
as the only potentially nonvanishing color chemical potential. This also
implies that the condition $\rho_3=0$ is trivially satisfied. Notice that
the fact that $\mu_3=0$, together with the ansatz chosen for
$\bar\Delta_a$, allow for a residual $r-g$ color symmetry, leading to
$\mu_r = \mu_g$.

The Matsubara sums can be now explicitly evaluated. One obtains
\begin{equation}
\Omega_{\rm MFA}(T) \ = \ - \, 2  \int
\frac{d^3p}{(2\pi)^3}\sum_{j=1}^6 \left[ 2\,T\; \ln \left(1 + e^{-\beta E_j}
\right)\; + \; E_j \right] \; + \; \frac{\bar\sigma^2}{2G}\; +
\; \frac{\bar\Delta^2}{2H} \; + \; {\cal U}(\Phi ,T) \ ,
\label{ommfa}
\end{equation}
where the quasiparticle energies $E_j$ are given by
\begin{eqnarray}
E_{1,2} & = & \varepsilon_p \mp \mu_b \\
E_{3,4} & = & \sqrt{(\varepsilon_p \mp \mu_r)^2 +\bar\Delta^2}- i\phi_3\\
E_{5,6} & = & \sqrt{(\varepsilon_p \mp \mu_g)^2 +\bar\Delta^2}+ i\phi_3~~.
\end{eqnarray}
Here we have defined $\varepsilon_p = \sqrt{|\vec{p}\, |^2 + M^2}$, being $M
= \bar m_0 + \bar\sigma$ the quark constituent mass.

The momentum integral is of course divergent. Following the usual NJL
regularization prescription, we cut the integrals of the zero-point
energies $E_j$ at a given cutoff $\Lambda$, while the logarithmic terms
are integrated up to infinity. In these last integrals, however, we assume
that NJL interactions vanish for energies above the cutoff, therefore we
set $\bar\sigma = \bar\Delta = 0$ for $|\vec p\, |>\Lambda\,$.

\section{Results}

\subsection{Non-superconducting case}

As in most of the effective models of QCD interactions, the presence of a
2SC phase shows up at low temperatures and moderate chemical potentials.
In order to study the effects of imposing the condition of color
neutrality, let us start by considering the NJL model without quark-quark
interactions ($H=0$), which means to set $\bar\Delta = 0$ in the effective
mean field thermodynamical potential in Eq.~(\ref{ommfa}). Notice that,
even in the absence of color superconductivity, one still may require a
nonvanishing color chemical potential $\mu_8$ owing to the presence of the
quark interactions with the Polyakov loop. According to the discussion in
the previous section, we take $\mu_3=0$, $\phi_8=0$, leading to
a real-valued mean field thermodynamical potential. Thus the set of six
relations (\ref{gap}-\ref{charges}) is reduced to only three nontrivial
equations; two for the order parameters
\begin{eqnarray}
\frac{\partial\, \Omega_{\rm MFA}}{(\partial\bar\sigma,\partial\phi_3)}
& = & 0 \ \ ,
\label{sigma}
\end{eqnarray}
and one for the constraint of color charge neutrality
\begin{eqnarray}
\frac{\partial\, \Omega_{\rm MFA}}{\partial\mu_8} = \rho_8 & = & 0 \ \ .
\label{rho8}
\end{eqnarray}

If we do not include quark-quark interactions, the model includes only
three free parameters, namely the quark current mass $\bar m_0$, the
coupling constant $G$ and the cutoff $\Lambda$. For definiteness, to
perform the numerical analysis we will use a standard set of values for
these parameters, taken from Refs.~\cite{Ratti:2005jh,Roessner:2006xn}:
\begin{equation}
\bar m_0 \ = \ 5.5 \ {\rm MeV} \ , \qquad
G \ = \ 10.1 \ {\rm GeV}^{-2} \ , \qquad
\Lambda \ = \ 650 \ {\rm MeV} \ .
\label{param}
\end{equation}

We start by considering the case of $T=0$. In this limit the Polyakov loop
decouples, since the mean field value $\phi_3$ can be removed from the
momentum integrals by a shift of the integration variable $p_4$. One gets
then two coupled equations, namely
\begin{eqnarray}
\frac{M - \bar m_0}{G} & = & \frac{M}{\pi^2} \int_0^\Lambda dp\
\frac{p^2}{E} \bigg[ \;2\,\sig(E + \mu_r) + 2\,\sig(E - \mu_r) +
\sig(E + \mu_b) + \sig(E - \mu_b) \;\bigg] ~,
\label{eqsigma} \\
0 & = & \frac{2}{\sqrt{3} \pi^2} \int_0^\Lambda dp\
p^2 \bigg[ \;\sig(E + \mu_r) - \sig(E - \mu_r)
- \sig(E + \mu_b) + \sig(E - \mu_b) \;\bigg] \ \ .
\label{eqmu8}
\end{eqnarray}
For low values of the chemical potential $\mu$, it is easy to see that
both equations are trivially satisfied for a wide range of values of
$\mu_8$, while from Eq.~(\ref{eqsigma}) the value of $M$ is fixed at $M =
M_0 = 324.1$ MeV. Thus, as expected, the system is in a phase in which the
chiral symmetry is strongly broken. These results are valid as long as
$\mu\leq M_0$. The allowed values of $\mu_8$ are those which satisfy
$|\mu_r|\leq M_0$ and $|\mu_b|\leq M_0$:
\begin{equation}
-\frac{M_0 -\mu}{2} \ \leq \ \frac{\mu_8}{\sqrt{3}} \ \leq \
\left\{
\begin{array}{lcl}
(M_0 +\mu)/2 & \ {\rm if} \ & \mu\leq M_0/3 \\
M_0 - \mu & {\rm if} & M_0/3 < \mu < M_0 \rule{0cm}{.5cm}
\end{array}
\right.
\end{equation}
Now if $\mu$ is increased above $M_0$ the value of the effective mass $M$
slightly decreases, while from Eq.~(\ref{eqmu8}) one has $\mu_8 = 0$. This
holds up to a critical value $\mu_c(T=0) = 342$ MeV. At this point one
finds a first order transition into a normal quark matter (NQM) phase, in
which the chiral symmetry is approximately restored. This behavior is
shown by the $T=0$ curve in the upper panel of Fig. 1, where we plot the
value of $\bar\sigma = M - \bar m_0$ as function of the chemical potential
$\mu$.

In order to determine the effect of the Polyakov loop one has to turn on
the temperature, solving Eqs.~(\ref{sigma}-\ref{rho8}). Let us first
consider the region of low temperatures ($T\ll\mu$) and values of $\mu$ up
to $\sim M_0$. In this region, while the value of the effective mass is
kept at $M\simeq M_0$, it is possible to show that the value of
$\phi_3$ is basically determined by the minimization of the effective
potential ${\cal U}(\Phi,T)$, which leads to $\phi_3\simeq 2\pi/3$
(or, equivalently, $\Phi\simeq 0$). In particular, it is interesting
to consider the range $0 \leq \mu \leq M_0/3$, where it is possible to
perform some analytic calculations to solve Eq.~(\ref{rho8}). From this
analysis, the value of $\mu_8$ is found to be given by the simple
expression
\begin{equation}
\frac{\mu_8}{\sqrt{3}} \ = \ 2\,\mu \; - \; T \ln{2} \ .
\label{linda}
\end{equation}
This result allows us to find the actual value of $\mu_8$ in the limit
$T=0$: one has $\mu_8/\sqrt{3} =2\mu$, which leads to $\mu_r=3\mu$,
$\mu_b=-3\mu$. Notice that the condition $\mu\leq M_0/3$ is necessary to
be in agreement with the constraints for $\mu_8$ that we established above
in the $T=0$ case. For low temperatures and $M_0/3\leq \mu\leq M_0$, it is
not easy to get a relation as simple as that in Eq.~(\ref{linda}).
However, it can be seen that in the  $T=0$ limit Eq.~(\ref{rho8}) implies
$\mu_r = M_0$, which means $\mu_8/\sqrt{3} = M_0 - \mu$. These results are
shown by the $T=0$ curve in the central panel of Fig. 1, where we quote
the value of $\mu_8$ as function of the chemical potential.

In this way, it has been shown that for nonzero temperatures and baryon
chemical potentials the interaction between the fermions and the Polyakov
loop requires the presence of a nonzero color chemical potential $\mu_8$
in order to keep color neutrality. This still holds in the $T=0$ limit if
one demands continuity in the values of $\mu_8$ and $\phi_3$. Hence
the fact that in this limit the Polyakov loop decouples does not mean that
the system is blind to the color symmetry breaking induced by
$\phi_3$.

For finite values of the temperature, the critical values of the chemical
potential $\mu_c(T)$ can be found numerically, defining a first order
phase transition line in the $T-\mu$ plane. This holds up to a given ``end
point'' $(\mu_E,T_E)$, which is found to be located at about (333 MeV, 64
MeV) for our parameter set. For $T > T_E$ the transition to the normal
quark matter phase proceeds as a smooth crossover. The corresponding
transition line can be established by looking at the peak in the chiral
susceptibility, defined by
\begin{equation}
\chi \ \equiv \ -\,\frac{1}{2}\,\frac{\partial^2 \Omega}{\partial
{\bar m_0}^{\; 2}} \ .
\end{equation}

\begin{figure}
\includegraphics[width=0.4\textwidth,angle=0]{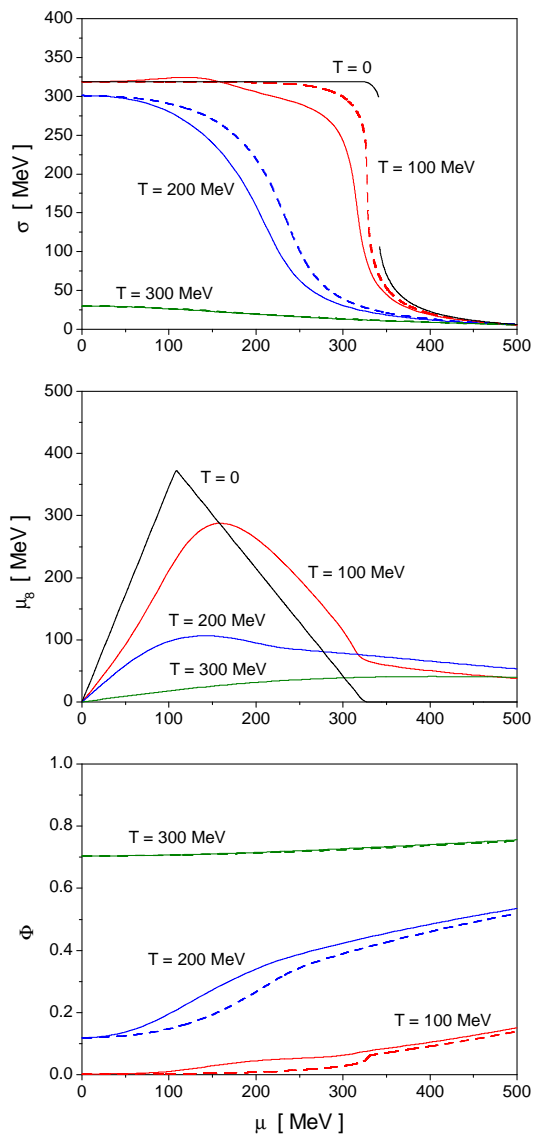}
\caption{Behavior of the effective mass $\bar\sigma$, the color chemical
potential $\mu_8$ and the traced Polyakov loop $\Phi$ as functions of
the chemical potential, for various values of the temperature, in the
non-superconducting two-flavor PNJL. Solid and dashed lines correspond to
the results with and without the imposition of color neutrality,
respectively. Notice that in the upper and lower plots the solid and
dashed curves corresponding to $T=300$~MeV are almost coincident.}
\label{fig1}
\end{figure}

Our numerical results are shown in Figs.\ 1 and 2. In Fig.\ 2 we represent
the $T-\mu$ phase diagram, denoting by solid and dashed lines the curves
corresponding to first order and crossover chiral restoration transitions
respectively. For comparison we also include the corresponding curves for
the PNJL model without the imposition of color neutrality. It can be seen
that the first order transition line extends up to about $T_E = 100$~MeV
in this latter case. Notice that at $\mu = 0$ the phase transition
temperature is found to be about 255 MeV, somewhat larger than the values
obtained from lattice QCD~\cite{Karsch:2006xs,Cheng:2006qk}. This is a
typical feature of local NJL models, and we do not adopt in our
calculations the rescaling procedure~\cite{Schaefer:2007pw} of the
Polyakov-loop potential which leads to a flavor-dependent lowering of the
critical temperature. It has been demonstrated that in nonlocal
generalizations of the PNJL model~\cite{Blaschke:2007np} the critical
temperature at $\mu = 0$ is in agreement with the lattice result. In
Fig.~1 we show the behavior of the effective mass $\bar\sigma$, the color
chemical potential $\mu_8$ and the traced Polyakov loop $\Phi$ as
functions of the chemical potential, for some representative values of the
temperature. The curves for $T=100$ MeV and $T=200$ MeV show clearly the
crossover transition, while for $T=300$ MeV the system is always in the
normal quark matter phase (see Fig.~2). For relatively low values of
$\mu$, the central and lower panels of Fig.\ 1 show the correlation
between the deconfinement transition and the restoration of color
symmetry: when the temperature is increased, the average mean field value
of the traced Polyakov loop rises from 0 (confinement) towards 1
(deconfinement); this comes together with a decrease of the color chemical
potential $\mu_8$, which has been introduced in order to recover color
neutrality. The effect is diluted in presence of a large baryon chemical
potential. Finally, in the upper and lower panels we have also included
with dashed lines the curves corresponding to the PNJL model without color
neutrality (in the central panel, these correspond simply to $\mu_8=0$).
The results for the models with and without color neutrality are
approximately coincident in the case of $T=300$ MeV.

\begin{figure}
\includegraphics[width=0.4\textwidth,angle=0]{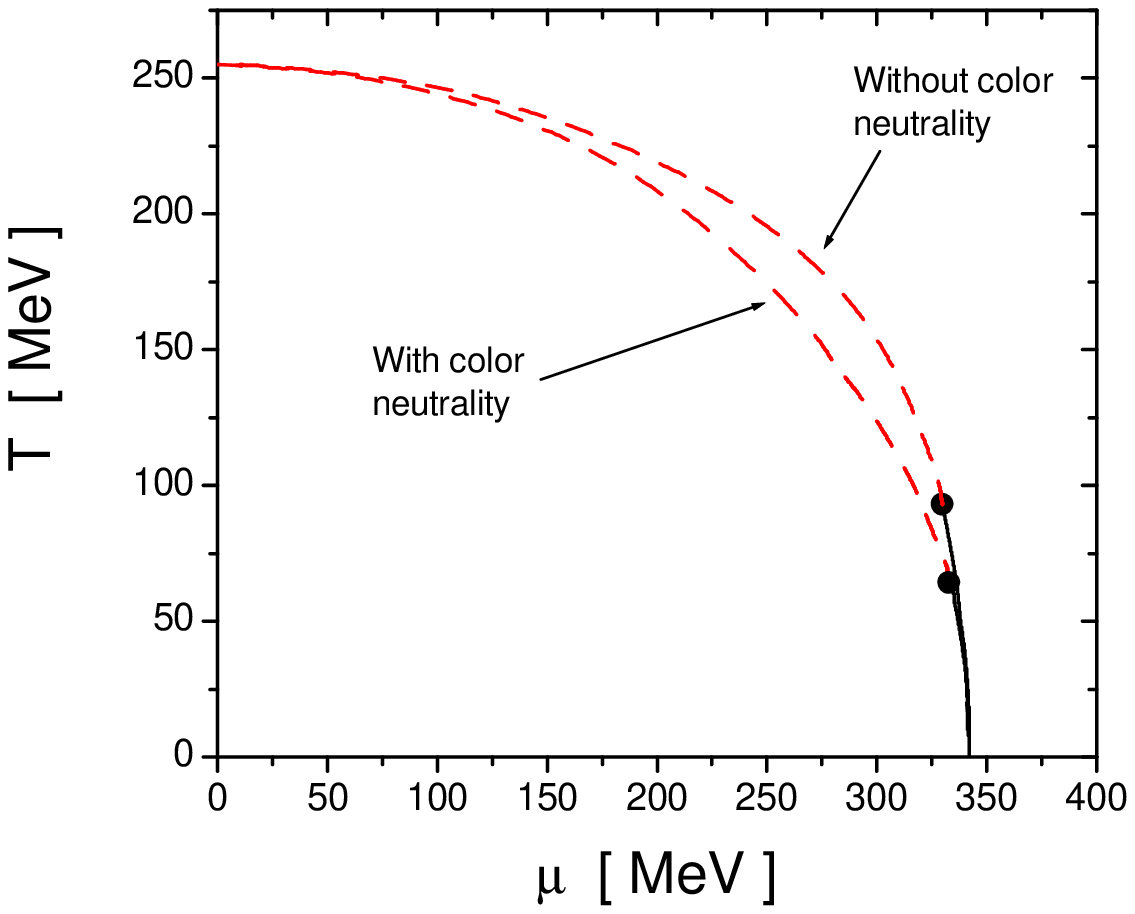}
\caption{Phase diagram for the non-superconducting two-flavor PNJL model
with and without the imposition of color neutrality. Solid and dashed
lines correspond to first order phase transition and smooth crossover,
respectively. The fat dots denote the position of the end points.}
\label{fig2}
\end{figure}

\subsection{Color-superconducting model}

Let us now take into account the quark-quark pairing interaction in
Eq.~(\ref{se}), with the ansatz $\bar \Delta_5 =\bar \Delta_7=0$, $\bar
\Delta_2 = \bar \Delta$. According to the discussion in Sect.\ II, we have
to solve the system of equations (\ref{sigma}), together with a fourth
equation $\partial\Omega_{\rm MFA}/\partial\bar\Delta = 0$. To proceed
with the corresponding numerical calculations one has to fix the value of
$H$, which cannot be directly obtained from $T=\mu=0$ phenomenology. For
definiteness we will consider values of the ratio $H/G$ around 0.75, which
is the value obtained from a Fierz rearrangement of the local quark-quark
interaction in Eq.~(\ref{se}).

\begin{figure}
\includegraphics[width=0.9\textwidth,angle=0]{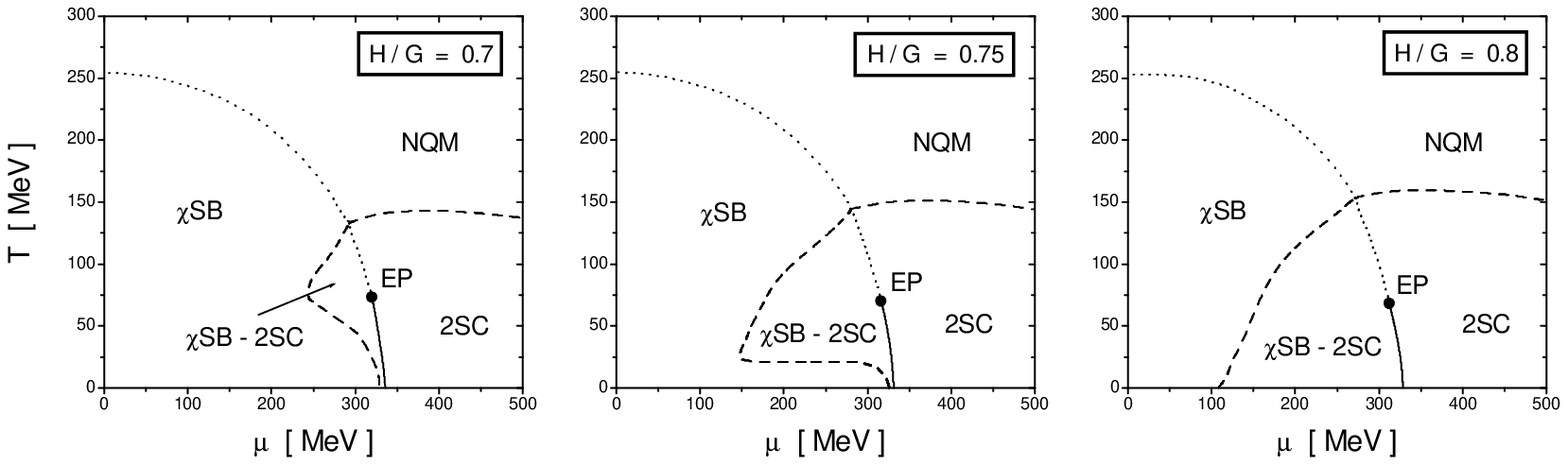}
\caption{$T-\mu$ phase diagrams for the neutral superconducting two-flavor
PNJL model, for various values of the ratio $H/G$. Solid, dashed and
dotted lines denote first order, second order and crossover phase
transitions, respectively. The fat dot indicates the end point.}
\label{fig3}
\end{figure}

\begin{figure}
\includegraphics[width=0.7\textwidth,angle=0]{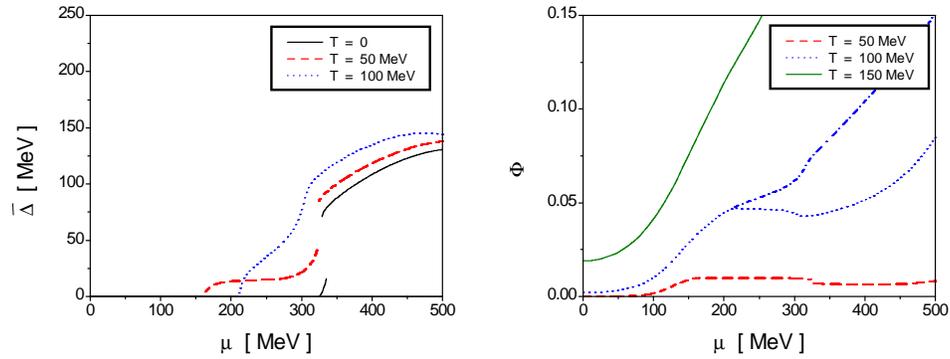}
\caption{Behavior of the color superconducting condensate $\bar\Delta$
(left) and the traced Polyakov loop $\Phi$ (right) as functions of the
chemical potential, for various values of the temperature and a coupling
ratio $H/G=0.75$. For comparison, the dashed-dotted line in the right
panel shows the curve corresponding to $T=100$~MeV in the
non-superconducting model ($H=0$).} \label{fig4}
\end{figure}

Our main results are shown in Figs.~\ref{fig3} and \ref{fig4}. In
Fig.~\ref{fig3} we show the arising $T-\mu$ phase diagrams. In order to
see the qualitative dependence of the results on the value of $H$, we have
considered the cases $H/G = 0.7$, 0.75 and 0.8. It is seen that now the
phase diagrams include in general a large region of two-flavor
superconducting (2SC) phases. For large values of the chemical potential
and relatively low temperatures, the system is in a 2SC phase in which the
chiral symmetry is approximately restored. Then, at $T \simeq 150$~MeV one
finds a second order phase transition into a normal quark matter (NQM)
phase. In addition, for intermediate values of $\mu$, the system gets into
a 2SC phase in which the chiral symmetry is still strongly broken
($\chi$SB). Here the quarks acquire large dynamical masses, which are not
substantially different from those obtained in the non-superconducting
model. The size of this phase region is large for $H/G \geq 0.75$, and
becomes reduced if the ratio is decreased. Once again, one finds a second
order phase transition into the non-2SC phase, while the chiral
restoration is driven by a first order phase transition for low
temperatures, and by a smooth crossover for temperatures above a given end
point. The shrinking of the $\chi$SB-2SC coexisting phase appears to be
the main qualitative effect of the change in $H/G$ within the considered
range.

In Fig.~\ref{fig4} we show the behavior of the mean field diquark
condensate $\bar \Delta$ (left panel) and the traced Polyakov loop $\Phi$
(right panel) as functions of the chemical potential for $T = 0$, 50~MeV,
100~MeV and 150 MeV. The curves correspond to the case $H/G = 0.75$. Both
the first order and the crossover chiral restoration transitions can be
noticed through their effects in the diquark condensate, while the size of
the diquark gap is found to be of the order of a hundred MeV for the bulk
of the temperature range (it decreases rapidly to zero near the phase
border, at $T\sim 150$ MeV). Concerning the traced Polyakov loop, it is
seen that the behavior is similar to that found in the case of the
non-superconducting model, in the sense that $\Phi \alt 0.15$ for
temperatures below 100 MeV and chemical potentials up to 500 MeV (see
lower panel of Fig.~1). Here the rise with $\mu$ is even less pronounced
than in the non-superconducting case, since one finds a plateau in the
$\chi$SB-2SC coexistence region (see curves corresponding to $T = 50$ and
100 MeV). For comparison, the behavior of $\Phi$ in the
non-superconducting model for $T=100$~MeV has also been plotted in the
figure. For $T\geq 150$ MeV there is no 2SC phase, thus there is no
difference between the behavior of $\Phi$ in both models. Finally, the
behavior of the color chemical potential $\mu_8$ in both superconducting
and non-superconducting models is also found to be qualitatively similar
(see central panel of Fig.~\ref{fig1}).

It has been noted earlier by Fukushima \cite{Fukushima:2008wg} that the
phase diagram of the PNJL model exhibits at densities $\mu>\mu_c$ and low
temperatures a so-called quarkyonic phase characterized by the coexistence
of confinement ($\Phi\ll 1$) and chiral symmetry restoration (see
also~\cite{Abuki:2008nm}). The possibility of such a phase was suggested
in Ref.~\cite{McLerran:2007qj} for large $N_c$ QCD where also the term was
coined. In a different context it has been discussed in
Ref.~\cite{Glozman:2008kn}. It is at present an open question whether a
quarkyonic phase shall exist in the real world with $N_c=3$. As a result
of the present study, we find that the requirement of color neutrality
together with the occurrence of diquark condensation are consistent with
the existence of a quarkyonic phase in the PNJL model. Moreover, as it is
shown in the right panel of Fig.~\ref{fig4}, the presence of diquark
condensation leads to a reduction of the expectation value of the traced
Polyakov loop $\Phi$ and thus enforce the confining aspects of the model.

It is important to point out that the presence of a nonvanishing Polyakov
loop variable $\phi_3$ breaks the color SU(3) symmetry down to SU(2), so
that the rotational invariance for the choice of the orientation of the
2SC gap under color neutrality constraints \cite{Buballa:2005bv} might be
lost. As a consequence, the above ansatz for $\bar\Delta_a$ may not
correspond to the true minimum of the thermodynamical potential which
instead might point to a different direction in the $\Delta_a$ space
\cite{Blaschke:2005km}. While accounting for such a disorientation of the
2SC condensate may lead to quantitative corrections, we believe that the
qualitative structure of the phase diagram for the PNJL model under color
neutrality constraints shown in the left panel of Fig.~\ref{fig3} remains
robust. The 2SC phase might coexist with chiral symmetry breaking and
confinement in the temperature range between about 50 MeV and 100 MeV at
densities below the onset of chiral restoration with possible observable
consequences in heavy-ion collision experiments, e.g., due to
diquark-antidiquark annihilation into lepton pairs \cite{Kunihiro:2007bx}.
Note that with the coexistence of $\chi$SB and 2SC phases the condition
for Bose-Einstein condensation (BEC) of diquark bound states is fulfilled
and the chiral restoration transition in the above temperature window may
therefore be characterized as a BEC-BCS crossover transition where the
diquark bound states undergo a Mott effect to unbound, but resonant
diquark Cooper pairs \cite{Sun:2007fc,Zablocki:2008sj}

\section{Conclusions}

In this paper we have explored the consequences of imposing the color
neutrality constraint within the framework of the standard mean field
treatment of the PNJL model at finite chemical potential. We have found
that the fulfillment of such a constraint leads to some changes in the
phase diagram of the model, as e.g.\ a shrinking of the $\chi$SB domain in
the $T-\mu$ plane and a shift of the end point to lower temperatures. The
most noticeable effect turns out to be the presence of a coexistence
region of $\chi$SB and 2SC phases, which may have consequences for
possible applications of the model in heavy-ion collision experiments.

The effects can be understood due to the nonvanishing  color chemical
potential $\mu_8$, which is introduced to compensate the color charge
$\rho_8$ induced by the color background field $\phi_3$. For a given
baryochemical potential $\mu$, $\mu_8 > 0$ reduces the chemical potential
of blue quarks, while simultaneously increasing that of red and green
ones. The net effect is a shift of the $\chi$SB phase border to lower
$\mu$ values than without the color neutrality constraint, due to the
nonlinear dependence of the quark mass gap on the chemical potential close
to its critical value for a given, not too low temperature. It is seen
that this effect is relatively slight, taking into account the theoretical
uncertainty that can be in general expected from an effective quark model.
In addition, the increase of chemical potentials of red and green quarks,
which pair in the 2SC channel, entails an effective lowering of the
critical value of $\mu=\mu_c$ for the onset of the 2SC phase, again for
not too low temperatures. One finds in this way a $\chi$SB-2SC coexisting
phase, whose size depends on the value of the parameter $H$ that drives
the quark-quark interaction. We notice that the 2SC border has been
obtained after choosing a given ansatz for the mean field values
$\bar\Delta_a$. The actual orientation of the 2SC gap in color space under
the presence of the Polyakov loop is a subject that deserves further
study. Moreover, it is worth to point out that at finite densities there
is no fundamental need to be restricted to Lorentz invariant condensates,
thus more general mean field configurations could be considered.

Signatures of the coexisting $\chi$SB and 2SC phases might become apparent
in not too energetic heavy-ion collision experiments (about 3-5
GeV/nucleon), as they are possible at the future facilities FAIR-CBM
(Darmstadt), NICA-MPD (Dubna) and J-PARC (Japan). Such a mixed phase can
give rise, e.g., to an enhancement of low-mass dilepton production due to
diquark-antidiquark annihilation. It is a realization of the BEC-BCS
crossover in quark matter. While such a transition has previously been
discussed in the NJL model by artificially increasing the diquark coupling
to large values~\cite{Zablocki:2008sj,Huang:2002,Kitazawa:2007}, which can
result in conflicts with the nuclear matter ground state, in the present
PNJL model the $\chi$SB-2SC coexisting phase has been obtained without any
artificial tuning of parameters. It just occurs as a consequence of color
neutrality in a domain of temperatures and chemical potentials around the
critical end point of the first order chiral phase transitions for the
standard Fierz-value of the diquark coupling, $H/G=0.75$, without
affecting the nuclear matter ground state at $T=0$.

Finally, it has been shown that the coexistence of confinement and chiral
symmetry restoration characterizing a quarkyonic phase in the PNJL model
at finite densities and low temperatures gets even reinforced due to color
superconductivity. The occurrence of diquark condensation reduces the
expectation value of the traced Polyakov loop and thus acts towards
strengthening the confinement aspect of the model.

\section*{Acknowledgments}

This work has been supported in part by CONICET and ANPCyT (Argentina),
under grants PIP 6009,  PIP 6084 and PICT04-03-25374. D.B. acknowledges
support from the Polish Ministry of Science and Higher Education under
contract No. N N202 0953 33. He is grateful for the hospitality of the
Institute for Nuclear Theory at the University of Washington and for
partial support from the Department of Energy during the completion of
this work.

\end{document}